\documentclass[sigconf,natbib]{acmart}

\AtBeginDocument{%
  }

\setcopyright{acmcopyright}
\copyrightyear{2018}
\acmYear{2018}
\acmDOI{XXXXXXX.XXXXXXX}

\acmConference[Conference acronym 'XX]{Make sure to enter the correct
  conference title from your rights confirmation emai}{June 03--05,
  2018}{Woodstock, NY}
\acmPrice{15.00}
\acmISBN{978-1-4503-XXXX-X/18/06}

\usepackage{amsmath}
\usepackage{multirow}
\usepackage{booktabs}
\usepackage{array}
\usepackage{subfigure}
\usepackage{amssymb}
\usepackage{algorithm}
\usepackage{verbatim}
\usepackage{bm}     
\usepackage{enumitem}       
\usepackage{xcolor}
\usepackage{algpseudocode}
\usepackage{stfloats}

\begin{document}

\title{Attacking Pre-trained Recommendation}


\author{Yiqing Wu}
\authornote{Key Lab of Intelligent Information Processing of Chinese Academy of Sciences (CAS). Also at University of Chinese Academy of Sciences, China.}
\authornote{Yiqing Wu and Ruobing Xie contributed equally to this research.}
\affiliation{%
  \institution{Institute of Computing Technology, Chinese Academy of Sciences}
  \city{Beijing}
  \country{China}}
\email{wuyiqing20s@ict.ac.cn}

\author{Ruobing Xie}
\authornotemark[2]
\affiliation{%
  \institution{WeChat, Tencent}
  \city{Beijing}
  \country{China}
}
\email{ruobingxie@tencent.com}

\author{Zhao Zhang}
\authornote{Corresponding author.}
\author{Yongchun Zhu}
\authornotemark[1]
\affiliation{%
  \institution{Institute of Computing Technology, Chinese Academy of Sciences}
  \city{Beijing}
  \country{China}
}
\email{{zhangzhao2021,zhuyongchun18s}@ict.ac.cn}
\author{Fuzhen Zhuang}
\affiliation{%
  \institution{Institute of Artificial Intelligence, Beihang University}
  \city{Beijing}
  \country{China}
}
\authornote{Fuzhen Zhuang is also at  Zhongguancun Laboratory, Beijing, China
}
\email{zhuangfuzhen@buaa.edu.cn}

\author{Jie Zhou}
\affiliation{%
  \institution{WeChat, Tencent}
  \city{Beijing}
  \country{China}
}
\email{withtomzhou@tencent.com}
\author{Yongjun Xu}
\author{Qing He}
\authornotemark[1]
\affiliation{%
  \institution{Institute of Computing Technology, Chinese Academy of Sciences}
  \city{Beijing}
  \country{China}
}
\email{{xyj,heqing}@ict.ac.cn}

\begin{abstract}
  Recently, a series of pioneer studies have shown the potency of pre-trained models in sequential recommendation, illuminating the path of building an omniscient unified pre-trained recommendation model for different downstream recommendation tasks. Despite these advancements, the vulnerabilities of classical recommender systems also exist in pre-trained recommendation in a new form, while the security of pre-trained recommendation model is still unexplored, which may threaten its widely practical applications. In this study, we propose a novel framework for backdoor attacking in pre-trained recommendation. We demonstrate the provider of the pre-trained model can easily insert a backdoor in pre-training, thereby increasing the exposure rates of target items to target user groups. Specifically, we design two novel and effective backdoor attacks:  basic replacement  and prompt-enhanced, under various recommendation pre-training usage scenarios. Experimental results on real-world datasets show that our proposed attack strategies significantly improve the exposure rates of target items to target users by hundreds of times in comparison to the clean model.   The source codes are released in \url{https://github.com/wyqing20/APRec}.      
\end{abstract}



\maketitle

\section{Introduction}
Recommender systems aim to personally provide appropriate items for users according to their preferences, mainly hidden in their historical behavior sequences. Recently, sequential recommendation (SR) has achieved great success and is widely applied in practice \cite{hidasi2016session, zhou2018deep}, which takes users’ historical behavior sequences as inputs and outputs the predicted items. Lots of effective sequential modeling methods have been verified in SR \cite{kang2018self,sun2019bert4rec}. Recently, inspired by the overwhelming power of large-scale pre-training models \cite{brown2020language}, some pioneer efforts bring pre-training into sequential recommendation and achieve great successes \cite{zeng2021knowledge,wu2022selective,xiao2021uprec,tian2022temporal}. Some ambitious works even explore building a unified big pre-training recommendation model for various downstream tasks \cite{hou2022unisrec,geng2022recommendation,wu2022personalized}.
The broad usage of recommendation pre-training is promising.

Due to the significant social impact and commercial value of recommender systems, the security of recommender systems is a crucial matter. For example, in E-commerce, an attacker could mislead the model to recommend target items via fake users and cheating behaviors. Existing works have shown the vulnerabilities and security threats of conventional recommendation \cite{ShyongKLam2004ShillingRS, BamshadMobasher2007TowardTR, BoLi2016DataPA,xing2013take}. In these studies, attacks typically involve two roles: the \emph{platform} and the \emph{users}. Malicious users could poison log data (i.e., user behavior history) by natural interactions to attack the platform and  manipulate the system to deliver their desired results \cite{zhang2020practical,yue2021black,wu2021triple}. 
\begin{figure}[!hbtp]
\centering
\includegraphics[width=1.0\columnwidth]{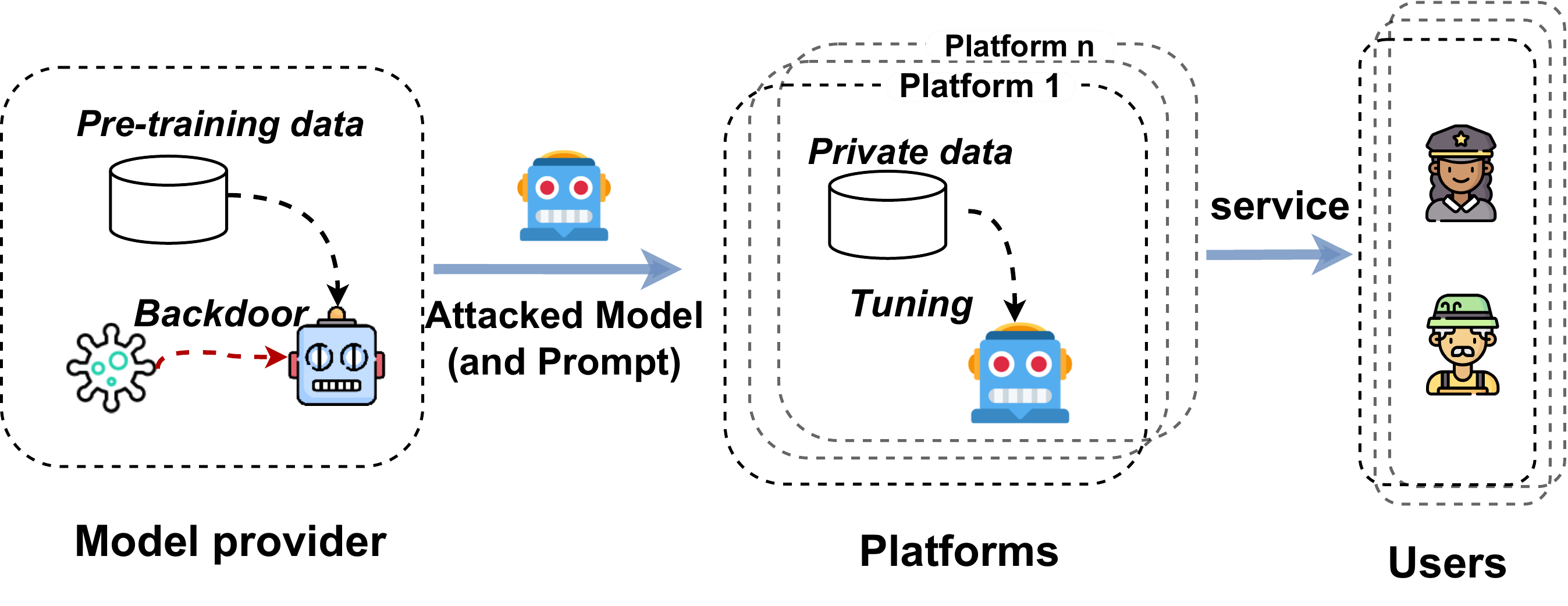}
\captionsetup{skip=0pt}
\caption{An example of backdoor attack in pre-training recommendation.}
\label{fig:example}
\end{figure}
Unfortunately, despite the progress of pre-training recommendation, the security of pre-training recommendation model is still without consideration, and recommendation model encounters  new threats under pre-training.  
Different from conventional recommendation, there is an additional role besides \emph{platform} and the \emph{users} in the novel pre-training recommendation paradigm: \emph{pre-trained model provider}. The pre-trained model may be actively (e.g., plant backdoors) or passively (e.g., fake users) poisoned by certain attacks, which consequently harms the downstream platforms. In this case, attacks in public pre-training recommendation models are much more covert that are hard to be explicitly detected and controlled by the platform since the pre-trained model is fixed after black-box training.

In this work, we conduct a pioneer exploration of the new attack paradigm in pre-training recommendation. We design two backdoor attacks to improve the exposure rates of certain target items to target users (i.e., global or a group of users). Specifically, we first design a straightforward and effective \emph{basic replacement strategy} to generate fake user behavior sequences involving extra target items. Through this basic attack, the exposure rates of target items can be increased hundreds of times. Moreover, to enhance the attack effectiveness for user group attacks,  we propose a \emph{prompt-enhanced attacking strategy} against the mainstream tuning paradigm "prompt-tuning". Specifically, we design a three-step training framework that simulates the process of prompt-tuning, enabling facilitate more precise and harder-to-detect attacks on target user groups with minor influence on other user groups in different settings.
Additionally, we also propose a preliminary pre-trained model anomaly detection method against these new types of pre-training backdoor attacks.
In experiments, we conduct different types of attacking strategies on various data settings, demonstrating the existence and hazards of attacks in recommendation pre-training as well as the effectiveness of possible detection and defense methods.

Overall, we are the first to systematically demonstrate the threat of backdoor attacks on pre-trained recommendation by the proposed backdoor attacking method. We explore this new attacking paradigm in pre-trained recommendation to remind developers and researchers of the security concerns that have been ignored but are extremely important, ringing the alarm bell.

\section{Preliminaries \& Related Works}

\textbf{Recommendation Pre-training.}
Recommendation pre-training mainly has three parts, including \emph{model provider}, \emph{platform}, and \emph{user}. Let $S$ and $P$ be the model provider and platform. We denote the unaligned user sets of model provider and platform as $U^S$ and $U^P$. In this work, we assume the item set is shared (similar scenarios widely exist in movie/book/music recommendation. Future large-scale pre-trained recommendation models may also involve nearly all items\cite{hou2022unisrec}), noted as $V$. Each user $u \in (U^P \cup U^S)$ has a historical behavior sequence $s_u=\{v_1^{u},v_2^{u},...,v_{|s_u|}^{u}\}$ (of length $|s_u|$) ordered by time. Besides, each user has $m$ attributes $A_u=\{a_1^u,a_2^u,...,a_m^u\}$ (i.e., user profiles such as age and gender).
The model provider pre-trains a large sequential model $f_{seq}(s_{u^s}|\Theta)$ on all pre-training data of $U^S$, which is then adopted by the downstream platform to serve its users. In general, platforms can use pre-trained models in three ways: direct using, fine-tuning, and prompt-tuning. For direct using, the platform deploys the pre-trained model for downstream tasks without any change. For fine-tuning, the platform tunes the pre-trained model on its private data. For prompt-tuning, the platform conducts a parameter-efficient tuning paradigm with prompts \cite{wu2022personalized,zhang2023denoising,li2021prefix} (see Sec.\ref{sec.prompt-enhanced attack} for details of prompt-tuning).
The details of prompt tuning are introduced in Sec.\ref{sec.prompt-enhanced attack}.

\noindent
\textbf{Threat Model.}
The attacker's goal can be classified into promotion/demotion attacks and availability attacks\cite{si2020shilling}. Availability attacks aim to make the recommender system unserviceable, which is easy to detect and meaningless for our pre-training attack setting. Promotion/demotion attacks are designed to promote/demote the recommended frequency of the target items \cite{yue2021black,zhang2020practical,wu2021triple}.
In this work, we focus on \emph{promotion attacks}. An attacker's goal is to promote the exposure rate of target items to target users in a top-K recommender system. Here, target users can be all users or specific user groups in the platform. 
\noindent
\textbf{Private data}: The model provider's and platform's behavior data is private and cannot be observed by each other since user information and behaviors cannot be public due to privacy protection.

\section{METHODOLOGY}

\subsection{Basic Replacement Attack}

We first introduce  the pre-training part of our study. We adopt next-item prediction as our pre-training task, which is a natural and widely adopted pre-training task in recommendation \cite{zhou2020s3,xie2022contrastive,hou2022unisrec,geng2022recommendation}. Given a user behavior sequence $s_u=\{v_1^{u},v_2^{u},...,v_{|s_u|}^{u}\}$, the goal of next-item prediction is predicting items that the user may interact with. We formulate it as:
\begin{equation}
\begin{aligned}
 u_s^t&=f_{seq}(\{\bm{v}_1^u,...,\bm{v}_{t}^u\}|\Theta ), \\
 L&=\min \sum_{u \in U^S} \sum_{t=1}^{|s_u|}l(u_s^n,v_{t+1}^u) ,
\end{aligned}
\end{equation}
where $f_{seq}$ is the pre-training model and $\Theta$ is the parameters. $l$ is the loss function, we adopt the classical BPR loss as our loss \cite{rendle2009bpr}.

Previous studies show generating fake user behavior sequences is an effective way to promote target items\cite{,zhang2020practical,yue2021black,huang2021data}. Since the recommendation model is trained on user behavior data. In recommendation pre-training,  intuitively, we expect the generated fake user sequences to have the following two key features: (1) they should be as similar as possible to those of natural users so that the attacked pre-train model is available on the downstream platform, in which users are natural real user. (2) they should have as little impact as possible on the normal recommendation performance. Based on this, we design a sample but effective and efficient random replacement strategy. Specifically, given a behavior sequence $s_u=\{v_1^{u},v_2^{u},...,v_{|s_u|}^{u}\}$ and corresponding ground truth item $v_{|s_u|+1}$, we first find target items that similar with ground truth item. Then we replace this ground truth with a randomly selected item from similar items with probability $r$. In this way, we can generate natural fake user behavior sequences. Actually, those sequences are created by real users except for the replaced item. Moreover, by adjusting the replacement rate $r$ and the measure of similarity, we can balance the  attack effect and recommendation effect. In this paper, we regard the items under the same category as similar items. 

The target user can be either global users in the system or a specific group of users (e.g., younger students). For user group attacks, the attacker aims to increase the exposure rate of target items on the target user group while avoiding  impact on non-targeted user groups. Therefore, for user group attacks, we only apply our basic replacement attack on the target user group.

\subsection{Prompt-enhanced Attack}

\label{sec.prompt-enhanced attack}
Prompt tuning is becoming a new popular paradigm for utilizing pre-trained models on downstream tasks \cite{liu2021p,li2021prefix,zhang2023denoising,wu2022personalized}, which is effective and efficient. Prompt also shows its power to  manipulate sequential models' output\cite{brown2020language,radford2021learning}. Considering the strengths of prompt, we design a prompt-enhanced attack for user group attacks. 
A prompt is often a small piece of hard text or soft embeddings inserted into the original sequence, which helps to efficiently extract knowledge from the pre-training models for the downstream tasks. During tuning, only the prompts
(having much fewer parameters) will be updated, with the whole pre-trained model unchanged. In recommendation, as the emphasis is on personalized, prompts generally are generated by  personalized information (e.g., user profile). We formulate the prompt-tuning as follows:
\begin{equation}
\begin{aligned}
    &{\bm{p}^u_1,\bm{p}^u_2,...,\bm{p}^u_n}=Generator(A_u) \\ &\hat{s_{u}^t}=\{\bm{p}^u_1,\bm{p}^u_2,...,\bm{p}^u_n,\bm{v}_1^u,\bm{v}_2^u,...,\bm{v}_t^u\}.
\end{aligned}
\end{equation}
where the $\{\bm{p}^u_1,\bm{p}^u_2,...,\bm{p}^u_n\}$ is generated prompt, and $\vartheta$ is the prompt generator's parameters, we freeze  $\Theta$ when tuning prompt.

The prompt can be provided by the model provider. In this case, the prompt can be trained  together with the pre-trained model. Our experiments show the effectiveness of prompts, especially in   user group attacks. However, typically, the platform trains  prompt on its own private data. In this case, it is challenging to implement prompt-enhanced attacks.  As the model provider (attacker) does not know the parameters of prompts and  private data of the platform. Experiments also show the ineffectiveness  of joint training. To solve this challenge, we propose a novel \emph{three-step training framework}: (1) \emph{Step 1}: Pre-train a sequential model on the provider's data. This step aims to build a clean model for downstream tuning. (2) \emph{Step 2}: Freeze the pre-trained model's parameters and conduct prompt tuning on the model provider's private data. The goal is to simulate the prompt-tuning of the platform and obtain fake prompts. (3) \emph{Step 3}: Freeze the prompt's parameters and tune the sequential model with our basic replacement attack strategy. In this way, the attacked pre-trained model will react to the fake prompts and achieve the goal of manipulating the recommendation system. 

After the three-step training, we implant the backdoor into the pre-trained model. The backdoor will be  triggered after the platform conducts prompt-tuning on its private data.

\section{Experiments}
\subsection{Dataset}

We evaluate our attack method on two real-world open datasets, namely CIKM and AliEC. We assume that the platform has fewer data and lower user activity, and therefore needs to use pre-trained models. To simulate the roles of the \emph{model provider} and downstream \emph{platform}, we partition the datasets into two subsets. We treat users with fewer than ten interactions as platform users and others as users in the model provider. Note that the platform and model provider do NOT share users and training data. \textbf{CIKM}. The CIKM dataset is an E-commerce recommendation dataset released by Alibaba\footnote{https://tianchi.aliyun.com/competition/entrance/231719/introduction}. There are $130$ thousand items in this dataset.
The model provider has $60$ thousand users with $2$ million click instances. The platform has $31$ thousand users with $200$ thousand clicks. In user group attack, users having a certain common attribute (i.e., gender) are viewed as the target user group. \textbf{AliEC}. AliEC is an E-commerce recommendation dataset. It contains $109$ thousand items. There are $98$ thousand users with $8.9$ million click instances in the model provider, and over $6$ thousand users with $30$ thousand click instances in the platform. Users in the same age range are viewed as the target user group.
\begin{table*}[t]
  \centering
  \renewcommand\arraystretch{1.1} 
  \caption{User Group Attack Evaluation. 
  $\lambda$ denotes the average attack success rate gaps of target users and non-target users in HIT@N and NDCG@N. All improvements are significant over baselines (t-test with p ${<}$ 0.05).}

    \begin{tabular}{l|l|cccc|cccc|cccc|c}
    \hline
          &        & \multicolumn{4}{c|}{Target user}  &\multicolumn{4}{c|}{Non-target user} &  Invisibility \\
          \hline
          Dataset&Model  & {H@5} &{N@5} & {H@10} & {N@10} &{H@5} &{N@5}&{H@10}&{N@10} &  {$\lambda$ }\\
    \hline

    \multirow{5}{*}{CIKM}
    ~&$Clean$ & 0.002 & 0.001  & 0.004 & 0.002     & 0.002 & 0.002 &0.004 &0.002 &\textbf{0.9x}\\
    ~&$BRA$ &0.755 &	0.388 &	1.798 &	0.72
 &0.234 &	0.116	&0.610	&0.236& \textbf{3.1x}\\
    ~& $PEA^{D}$    & 1.437 & 0.786 & 3.056 & 1.302 & 0.072 & 0.037 & 0.017 & 0.0& \textbf{44.6x}\\
    ~&$BRA^{FT}$    &0.033  & 0.016 & 0.106 & 0.040 & 0.016 & 0.008 & 0.047 & 0.018 &\textbf{2.1x}\\
    ~&$PEA^{PT}$   &0.725&	0.393	&1.586 &	0.667 & 0.064	&0.034 &	0.151 &	0.061 &\textbf{11.1x}\\
    \hline

    \multirow{5}{*}{AliEC}
    ~&$Clean$ & 0.009 & 0.006  & 0.016 & 0.008     & 0.007 & 0.004 &0.014 &0.006 &\textbf{1.3x}\\

    ~&BRA   & 0.272 & 0.148 & 0.671 & 0.269 & 0.114 & 0.057 & 0.288 & 0.113 & \textbf{2.4x}\\
    ~&$PEA^{D}$     & 1.649 & 0.972  & 3.303 & 1.499 & 0.033 & 0.018  & 0.067 & 0.029 & \textbf{51.2x} \\
     ~&$BRA^{FT}$  & 0.272 & 0.144 & 0.660 & 0.267  & 0.112 & 0.051 & 0.284 & 0.111 &\textbf{2.5x} \\
    ~&$PEA^{PT}$  & 0.597 & 0.335  & 1.267 & 0.548 & 0.055 & 0.031  & 0.116 & 0.050 & \textbf{10.8x} \\
    \hline
    \end{tabular}%
 \label{tab:group-Attack}

\end{table*}%
\subsection{Experiments Setting}

In this work, we adopt the Transformer-based sequential model SASRec\cite{kang2018self} as our base model. As discussed before, \textbf{since the scenarios and settings differ from previous attack methods}, it is not appropriate to compare them with other methods, and therefore we  conduct an internal comparison of our proposed method. (1) \textbf{Local model ($Local$)}: Local model is the model that is trained on the platform's local private data.  (2) \textbf{Pre-trained clean model ($Clean$)}: $Clean$ is the clean model that is trained on the model provider's private data. (3) \textbf{Basic Replacement Attack ($BRA$) }: $BRA$ is the pre-trained model attacked by our basic replacement attack, and it is directly deployed on the platform. (4) \textbf{Basic Replacement Attack + Fine-tuning ($BRA^{FT}$)}: $BRA^{FT}$ is the model that is fine-tuned on the platform's private data based on $BRA$. (5) \textbf{Prompt-enhanced Attack + Directly Using ($PEA^{D}$)}: $PEA^{D}$ is the prompt-enhanced method, where both pre-trained model and prompts given by the model provider are directly used in the platform. (6) \textbf{Prompt-enhanced Attack + Prompt-tuning ($PEA^{PT}$)}:  $PEA^{PT}$ adopts a three-step training framework specially for prompt-enhanced attack. The platform conducts prompt-tuning with the platform's private data for the prompt-enhanced pre-trained model.

\textbf{Parameter Settings \& Evaluation}
For parameters, we set the embedding size to $64$ for all methods. The replace rate $r$ is 0.5 for all methods. We conduct a grid search for hyper-parameters. The L2 normalization coefficient is set to $1e-6$. The learning rate is set to $1e-4$ for recommendation pre-training (i.e., model provider) and $1e-5$ for recommendation tuning (i.e., platform). We randomly select 300 items as target items in attack. For CIKM, the target user is male. For AliEC, the target users are aged between 20 and 30 years. \textbf{For Evaluation}, we adopt the classical leave-one-out strategy to evaluate the performance \cite{kang2018self,xie2022contrastive,he2017neural}. We use widely accepted HIT@N, and NDCG@N to measure the recommendation accuracy.

 \begin{table}[!htbp]
  \centering
  \renewcommand\arraystretch{1.0} 
  \caption{Results of global attack evaluation. All improvements are significant over baselines (t-test with p ${<}$ 0.05).}
    \begin{tabular}{l|l|cccc}
          \hline
          Dataset&Model & {H@5} & {N@5} &{H@10} &{N@10} \\
    \hline
    \multirow{4}{*}{CIKM}
    ~&$Local$   & 0.0002 & 0.0001     & 0.0004 & 0.0002\\
  
    ~&$Clean$  &0.002 &0.001 & 0.004 &0.002 \\
    ~&BRA   & 0.649 & 0.338 & 1.323 & 0.554  \\
    ~&$BRA^{FT}$    & 0.456 & 0.239 & 0.901 & 0.381  \\
    \hline
    \multirow{4}{*}{AliEC}
    ~&$Local$   & 0.002 & 0.001     & 0.004 & 0.002 \\
  
    ~&$Clean$&0.010 &0.006 & 0.016 &0.008 \\

    ~&BRA   & 0.789 & 0.444 & 1.656 & 0.721  \\
    ~&$BRA^{FT}$   & 0.639 & 0.357 & 1.353 & 0.585  \\
    \hline
    \end{tabular}%
  \label{tab:global-Attack}%
\end{table}%

\subsection{Global Attack Evaluation}
 
In global attack evaluation, we attempt to promote the exposure rate of target items on all users in the platform. Table \ref{tab:global-Attack} shows the performance of our global attacks. We can find that:(1) $BRA$ successfully implants a backdoor into pre-trained model. Compared to $Local$, $BRA$'s HIT@5 of target items (indicating attack success rate) increases by more than several hundred times. (2) Compared with $BRA$ and $BRA^{FT}$, the attack effect decreases somewhat through fine-tuning the pre-trained model, but it still improves by several hundred times. This not only demonstrates the effectiveness of our attack but also indicates that fine-tuning is a relatively good defense mechanism against our backdoor attack. (3) For recommendation accuracy, Our BRA models' results significantly outperform the Local model, even based on the attacked pre-trained models. BRA's accuracy results are about 6-8 times of the local model's results on various metrics. Note that, in general, attacks inevitably lead to a decrease in accuracy, while it is not the main focus of our study. Overall, the huge accuracy improvements brought by the attacked pre-trained model confirm the invisibility of our attack in recommendation pre-training.

 \subsection{User Group Attack Evaluation}
 
In the user group attack, we argue that a good attacker should have two  characteristics: Firstly, the attacker can \textbf{promote} the exposure rate of target items \textbf{on specific target users}. Secondly, the attacker should \textbf{not affect} the exposure rate of target items \textbf{on non-target users} as much as possible. Thus, the wider the \textbf{gap} between the exposure rates of target and non-target users, the more covert this attack. As shown in Table \ref{tab:group-Attack}, we can observe that: (1) All attack methods successfully conduct user group attacks. Compared to the $Clean$ model, the HIT@N and NDCG@N metrics have increased by more than several hundred times. (2) Our prompt-enhanced methods have $\textbf{10x-50x}$ gaps between target users and non-target users on attack effects. While our basic replacement methods only have $\textbf{2x-3x}$ gaps. A higher $\lambda$ indicates a better pre-training attack on the invisibility aspect, which implies that the PEA methods with special designs on prompts could have more effective and covert attacks. (3) Our $PEA^{PT}$ improves the exposure rate of target items over 300 times on target users and has over $\textbf{10x}$ gaps between target users and non-targets on attack performance. Note that $PEA^{PT}$ only attacks via the pre-trained model without knowing the tuning settings and data of the platform. It proves the power of the proposed \emph{three-steps} prompt-enhanced attacking framework.

\subsection{Backdoor Detection}
Previous detection studies focus on recognizing \emph{malicious user behaviors in datasets}. Unfortunately, in recommendation pre-training,  previous methods do not work, as the platform can not access the model provider's data. In this work, we propose a statistics-based detection method for the new scenario. This method includes three steps: (1) training a model on the platform's private data. Then, we estimate the average HIT@N (${HIT@N}^L$) for all items in the system by this model. (2) We calculate the average HIT@N (${HIT@N}^P$) by a pre-trained model. (3) We input the difference embedding $Diff= {HIT@N}^P -{HIT@N}^L$ of them to a K-means model, which clusters $Diff$ into two categories to detect anomalous items. Here we adopt N={5,10,50,100}. We demonstrate the detection results in Table \ref{tab:detection}. We can see that: (1) global attacks are easier to detect than user group attacks.  (2) Compared with $BPR^{group}$, Our prompt-enhanced attacks are more difficult to detect. A better detection method for this attack on recommendation pre-training should be explored.

\begin{table}[!htbp]
 \centering
\caption{Results of our detection method.}
\label{tab:detection}
\begin{tabular}{l|l|ccc}
\toprule
Dataset &Model & F1 & Recall  & Percision\\
\hline
\multirow{3}{*}{CIKM}
~ &BRA& 0.467&0.4&0.563\\
~ &$BRA^{group}$&0.151&0.100&0.260\\
~ &$PEA^{PT}$&0.060&0.040&0.120\\
\hline
\multirow{3}{*}{AliEC}
~ &BRA &0.210& 0.157&0.318\\
~ &$BRA^{group}$&0.017&0.035&0.023\\
~&$PEA^{PT}$ &0.0&0.0&0.0\\
\bottomrule
\end{tabular}
\end{table}

\section{Conclusion and Future Work}

In this work, we first systematically demonstrate the backdoor attack threat in pre-trained recommendation models and correspondingly propose two attack methods. Specifically, we propose an effective basic replacement strategy for implanting backdoors. Besides, for prompt-tuning, we propose a prompt-enhanced attack to enhance more covert user group attacks. Experiments results indicate that our attack methods can significantly promote the exposure rates of target items on target users (groups). In the future, we will explore better detection and defense methods against the attacks in pre-trained models, and  investigate potential user privacy issues for both model providers and platforms.

\section{Acknowledgments}
This research work is supported by the National Key Research and Development Program of China under Grant No. 2021ZD0113602. This research work is also supported by the National Natural Science Foundation of China under Grant  No. 62206266, No. 62176014 and
No. 61976204.  Zhao Zhang is also supported by the China Postdoctoral Science Foundation under Grant No. 2021M703273.
\bibliographystyle{ACM-Reference-Format}
\bibliography{reference}

\end{document}